\documentclass[useAMS,usenatbib]{mn2e}
\usepackage{graphicx,amssymb,epsfig}

\title[$U\!BV\!(RI)_C$ photometry of open clusters]
 {$U\!BV\!(RI)_C$ photometry of the open clusters Be~15, Be~80, and NGC~2192}

\author [M. T. Tapia et al.]
{M. T. Tapia$^{1,2,3}$ \thanks{E-mail: mtt@iac.es(MTT),
schuster@astrosen.unam.mx(WJS)}, W.J. Schuster$^{2}$, R. Michel$^{2}$,
C. Chavarr\'{\i}a-K.$^{2}$, W. S. Dias$^{4}$,
\newauthor R. V\'azquez$^{2}$, and A. Moitinho$^{5}$.\\
$^{1}$ Instituto Astrof\'isico de Canarias, E-38200, La Laguna, Tenerife, Spain\\
$^{2}$ Instituto de Astronom\'{\i}a, Universidad Nacional Aut\'onoma de M\'exico,
Ensenada 22860, Baja California, M\'exico\\
$^{3}$ Universidad Aut\'onoma de Baja California, Ensenada 22860, Baja California, M\'exico\\
$^{4}$ UNIFEI, Instituto de Ci\^encias Exatas, Universidade Federal de Itajub\'a, Itajub\'a MG, Brazil\\
$^{5}$ SIM/IDL, Fac. de Ciencias, Universidade de Lisboa, Ed. C8, Campo Grande, 1749-016, Lisboa, Portugal
}
\begin{document}

\date{Accepted XXXX. Received YYYY; in original form ZZZZ}

\pagerange{\pageref{firstpage}--\pageref{lastpage}} \pubyear{2007}

\maketitle

\label{firstpage}

\begin{abstract}

The three open clusters Be~15, Be~80, and NGC~2192 have been observed using CCD
$U\!BV\!(RI)_C$ photometry at the San Pedro M\'artir Observatory, M\'exico within the
framework of our open-cluster survey.  The fundamental parameters of interstellar
reddening, distance, and age have been derived, and also the metallicity for
NGC~2192 (solar metallicity has been assumed for the other two).  By shifting the
colours of Schmidt-Kaler in the $(U$--$B, B$--$V$) two-colour diagram along the
appropriate reddening vector, the interstellar reddenings have been derived as
$E(B$--$V)= 0.23\pm0.03$ mag for Be~15, $1.31\pm0.05$ for Be~80, and
$0.16\pm0.03$ for NGC~2192.  Evidence is shown for a variable interstellar
extinction across the cluster Be~80.  For NGC~2192 a nicely consistent fit is
obtained for both the interstellar reddening and the metallicity ($[Fe/H] = -0.31$)
using simultaneously the F-type and red-clump stars.

By fitting isochrones to the observed sequences of these three clusters in various
colour--magnitude diagrams of different colour indices, ($B$--$V$, $V$--$I$, or
$V$--$R$) the averages of distance moduli and heliocentric distances
(($V$--$M_{V})_{o}$ (mag); $d$(pc)) are the following:  $(10.74\pm0.01$; 1202) for Be~15,
$(10.75\pm0.01$; 1413) for Be~80, and $(12.7\pm0.01$; 3467) for NGC~2192,
and the averages of the inferred best ages $(\log({\rm age})$; age (Gyr)) are
$(8.6\pm0.05$; 0.4) for Be~80, and $(9.15\pm0.05$; 1.4) for NGC~2192; for Be~15
there are two distinct possibilities for the age fit, depending on the membership
of three brighter stars, (9.35 or 9.95$\pm0.05$; 2.2 or 8.9 ).  The need for
spectroscopic observations in Be~15 is emphasized to select between alternate
reddening and age solutions, and for deeper $U\!BV$ observations in Be~80 to study
in greater detail the variable interstellar, or intracluster, reddening across
this cluster.

\end{abstract}


\section{Introduction}

Stellar clusters are ideal for research being groups of stars born
simultaneously, under the same physical conditions, situated at the same
distance, but with a wide range of stellar masses.  Altogether stellar clusters
are excellent probes for the study of the structure, and dynamical and chemical
evolutions of the Galaxy, since their interstellar reddenings, chemical
abundances, distances, and ages can be determined with precision using various
colour-colour and colour-magnitude diagrams from the $U\!BV\!(RI)_C$ photometry,
as well as stellar models and isochrones. In effect, the younger open stellar
clusters have been utilized as tracers of spiral structure of the Galaxy
(e.g.~Becker \& Fenkart 1970, Moffat \& Vogt 1973, Feinstein 1994), as tracers
of Galactic three-dimensional structure (e.g.~Cabrera-Ca\~no et al.~1990, Phelps
\& Janes 1994, Moitinho 2002), and in the analysis of the structure and chemical
evolution of the Galactic disc (e.g.~Janes 1979, Friel 1995, Piatti et
al.~1995, Twarog et al.~1997).  Individually, stellar clusters can be used to
constrain and test the theories of stellar formation and evolution.  The detailed
comparison between photometric diagrams and theoretical isochrones, calculated
from models of stellar evolution, has led to enormous advances in the understanding
of stellar structure and evolution, and is still providing very useful information
concerning the effects of convective overshooting, chemical compositions,
gravitational settling, and radiative accelerations (Meynet et al. 1993, Kassis
et al.~1997, Michaud et al.~2004).  These photometric diagrams are also useful
in searching for interesting and peculiar stars, such as Be stars, blue 
stragglers, recently formed stars, and the brightest red giants (Mermilloid
1982a,b; Mermilloid \& Mayor 1989; Chavarr\'{\i}a-K 1994; Moitinho et al.~2001).\\

In these Galactic studies, one of the more severe observational limitations is
due to the absence of photometric data for nearly half of the approximately
1500 open clusters known, combined with the lack of homogeneity in the
observations and analyses for those clusters which have been studied.  These
limitations form the crux of various controversies, such as:  (1) whether or
not there is a metallicity gradient as a function of the Galactocentric radius
(Janes 1979, Twarog et al. 1997, Carraro et al. 1998), (2) whether or not there
is a vertical metallicity gradient with respect to the Galactic disc (Piatti et
al.~1995), and (3) uncertainties in the characteristics of the Galactic 
structure, for example the spiral structure (i.e.~the number and positions of 
the various arms).\\

The catalogue of Lyng{\aa} (1987), which includes distances for 422 open clusters,
has constituted the observational basis for a large number of astronomical studies,
has led to important conclusions about the Galactic disc, and has been very useful
for planning subsequent observations by other astronomers.  However, this catalogue
has been built from parameters obtained by various authors, with diverse observing
techniques, distinct calibrations, and different criteria for determining the
stellar ages, rendering it very inhomogeneous and limited for studies requiring
precision in the measurement of these fundamental parameters.  As an example of the
precision and accuracy that one can expect due to the effects of these
inhomogeneities, we refer to Janes \& Adler (1982), who found that distance moduli
obtained by two or more authors have a mean difference of $0.55$ mag (see also
WEBDA:  http://www.univie.ac.at/webda/navigation.html).\\

To remedy these inaccuracies, a project (`A $U\!BV\!(RI)_C$ survey for open
clusters of the northern hemisphere') has been developed whose principal
objective is to obtain and to analyze in a systematic way a set of CCD data,
taken with the $U\!BV\!(RI)_C$ photometric system, for the more than 300 open
clusters visible from the National Astronomical Observatory, San Pedro M\'artir
(SPM), most of which have been previously unstudied (Schuster et al.~2007;
Michel et al.~in preparation).  With this photometry, which is homogeneous regarding
instrumentation, observing techniques, reduction methods, and analysis, the
following will be obtained:
(a) a uniform, photometric reference system via open clusters to which other
photometric studies can be transformed,
(b) an atlas of colour-colour (CC) and colour-magnitude (CM) diagrams for Galactic
open clusters with distinct fundamental parameters,
(c) a homogeneous data set of interstellar reddenings, metallicities, distances,
and ages for a large set of Galactic open clusters,
(d) an increased number of old and distant open clusters.
(e) selection criteria lists for subsequent spectroscopic and deeper photometric
studies of open clusters.\\

\section[]{The selection of the clusters}

The open clusters for our `quick' survey have been selected from the large
(and mostly complete) catalogue of Dias et al.~(2002), `Optically visible open
clusters and Candidates', which is now also available and updated through 2007
at the CDS (Centre de Donn\'ees Astronomiques de Strasbourg) as
catalogue VII/229 (or B/ocl).  Our subset of clusters are those observable from
SPM (latitude $+31\degr 44'$) and which have little or no previous studies, such
as Be~15 and Be~80 of this paper; a few others which have been analyzed earlier
have also been added for comparison purposes, such as NGC~2192 (Tapia-Peralta
2007).  More details concerning the selection of clusters and the project
motivation will be presented in the principal, introductory publication of this
survey (Michel et al.~in preparation).  The three open clusters of this paper, Be~15,
Be~80, and NGC~2192 represent a small, carefully selected sample used to test
and to define the analysis procedures of this `quick' survey; these clusters
have been well observed, provide probable turn-offs in the G-, B-, and F-type
stars, respectively, and include small to moderately large interstellar reddenings.

\section[]{CCD observations and reductions}

A CCD $U\!BV\!(RI)_C$ survey of northern open clusters has been undertaken at SPM
using always the same instrumental setup (telescope, CCD, filters), observing
procedures, reduction methods, and system of standard stars (Landolt 1983, 1992).
The CCD $U\!BV\!(RI)_C$ observations of this paper have been made exclusively with
the 0.84-m f/13 Cassegrain telescope at SPM, during the night of 2001 June 25/26
for the open cluster Be 80, and the nights 2002 February 6/7 and 8/9 for the
clusters NGC 2192 and Be 15, respectively.  The telescope hosted the filter-wheel
`Mexman' with the SITe~1 (SI003) CCD camera, which has a 1024$\times$1024 square
pixel array, with a pixel size of $24\mu$m$\times24\mu$m; this CCD has non-linearities
less than 0.45 per cent over a wide range, no evidence for fringing even in the $I$ band, and
Metachrome II and VISAR coverings to increase sensitivity at the blue and
near-ultraviolet wavelengths.  This telescope was re-focused before the observation
of each open cluster, usually using the $V$ filter of our parfocal set of $U\!BV\!(RI)_C$
filters.  The open clusters have been observed with exposure times of $3 \times 240$
seconds for the $U$ filter, $3 \times 180$ for $B$, $3 \times 100$ for $V$, $3 \times 100$
for $R$, and $3 \times 120$ for $I$; for Be~15 and NGC~2192 extra exposures in the filter
$U$ were made to improve the signal-to-noise.  Also, for Be~80 exposures as short as
10 seconds have been made in the $R$ and $I$ filters to de-saturate the brightest stars
of the field, and as short as 25 seconds to de-saturate the brightest stars of Be~15
and NGC~2192.

Each night several standard-star fields from Landolt (1992) were observed to permit
the derivation of the photometric transformations to the system of Johnson-Cousins
and the coefficients of the atmospheric extinction.  For the June 2001 observing run,
six Landolt groups were used, containing 23 different standard stars:  PG1633+099
(5 stars), MARK~A (4 stars), PG1528+062 (3 stars), PG1530+057 (3 stars), PG1525-071
(4 stars), and PG1657+078 (4 stars), with a range in $(B$--$V)$ from $-0.252$ to
$+1.134$, in $(U$--$B)$ from $-1.091$ to $+1.138$, and in $(V$--$I)$ from $-0.296$ to
$+1.138$.  Thirteen to 25 observations of these Landolt standards were made per night.
For the February 2002 run eight Landolt groups were employed, containing
35 different standard stars:  PG1047+003 (4 stars), PG1323-086 (5 stars), SA95
(5 stars), SA107 (4 stars), PG0942-029 (5 stars), SA104 (4 stars), PG0918+029
(5 stars), and PG1528+062 (3 stars), with a range in $(B$--$V)$ from $-0.294$ to
$+1.412$, in $(U$--$B)$ from $-1.175$ to $+1.265$, and in $(V$--$I)$ from $-0.280$ to
$+1.761$.  Fifty-two to 72 observations of these Landolt standards were made per night,
except one night cut short by clouds, when only 15 observations were managed.  The
standard-star fields have been observed with exposures of $1 \times 240$ seconds for
the $U$ filter, $1 \times 120$ for B, $1 \times 60$ for V, $1 \times 60$ for R, and
$1 \times 60$ for I.

Usually one, or more, Landolt fields were re-observed with an air-mass range of at
least 0.70 in order to measure the coefficients of the atmospheric extinction.  For
example, the night when Be~80 was observed, an approximate air-mass range of 1.29
was achieved, and a range of 1.26 for NGC~2192.  Due to the wide band-passes of
these Johnson-Cousins filters, second-order colour terms were included in these
atmospheric-extinction corrections.  For the large air-mass observations required
for an atmospheric-extinction determination, the filters were frequently observed
with both forward and backward sequences (i.e.~$U\!BV\!RIIRV\!BU$); this was
occasionally done also for other standard-star fields to increase the accuracy,
precision, and observing efficiency of the photometric observations.

The usual calibration procedures for CCD photometry have been carried out during
each of our observing runs.  Fifty to a hundred `bias' exposures have been made each
night, and fifty or more `darks' during each run with exposures of 4--15 minutes,
according to the longest of our stellar exposures;  these `darks' were usually made
during the non-photometric nights.  Flat fields were obtained at the beginning and
end of the nights by observing a clear patch of sky; at least 5 flat fields were
obtained for each filter per night with exposures greater than 5 seconds, and with
small offsets on the sky between each flat-field exposure.

The data reductions of this CCD photometry have been carried out using the usual
techniques and packages of IRAF.\footnote[1]{IRAF is distributed by NOAO, which is
operated by the Association of Universities for Research in Astronomy, Inc.,
under cooperative agreement with the NSF}  Aperature photometry and PSF photometry
have been used for handling the standard-star and cluster-star observations,
respectively (Howell 1989, 1990; Stetson 1987, 1990).  More details concerning the
instrumentation and the observing and reduction procedures of this project will be
given in an introductory paper (Michel et al.~in preparation).

For the analyses of these open clusters, an awk macro, `elipse', has been developed
which allows us to concentrate on the centre of a cluster, as defined by visual
inspection in a visual ($V$) or red ($R$) image, helping to increase the contrast of
the cluster with respect to the surrounding fields in the various colour-colour (CC)
and colour-magnitude (CM) diagrams.  This macro defines an elliptical area on the image,
and when applied to the CCD data file, retains all stars within this elliptical area
for further analyses and excludes all stars outside this area.  The input parameters
of this macro are the centre of the ellipse (X,Y coordinates in pixels), the major and
minor axes of the ellipse in pixels, and an angle of rotation in degrees.  This macro
has been used interactively by selecting the input parameters, then plotting the
resulting elliptical field, and finally iterating to a more concentrated central
region as seen in the visual or red image.  This ellipse has been fit as tightly as
possible to each cluster while still including all obvious visible members.  This process
is somewhat subjective, and has been applied prior to the plotting of any CC or CM
diagrams.

A java-based computer program (`SAFE' by J.~McFarland, 2009) has also been
utilized for the visualization and analysis of the photometric data of these
open clusters (Schuster et al.~2007). This system is capable of displaying
each cluster's data simultaneously in different CC and CM diagrams and has an
interactive way to identify a star, or group of stars, in one diagram and to
see were it falls in the other diagrams, thus facilitating the elimination of
field stars and the apperception of cluster features.  This program is capable
of displaying up to 16 different diagrams for one cluster while processing up to
20 clusters at the same time.

X and Y positions (in pixels), and standard $U\!BV\!RI$ CCD photometry and observing
errors for the open clusters Be~15, Be~80 and NGC~2192 are presented in Tables~1--3,
respectively.

\begin{table*}
\begin{minipage}{185mm}
\caption{Standard $U\!BV\!RI$ CCD photometry and observing errors for the open cluster Be~15.
The columns give successively the following:  X and Y (pixels), the position of a star in the CCD
field; $U, B, V, R$, and $I$ (magnitudes), the standard photometry of this star; and $\sigma_{U},
\sigma_{B}, \sigma_{V}, \sigma_{R}$, and $\sigma_{I}$ (magnitudes), photometric errors as provided
by IRAF.  Values of `99.999' indicate photometry that was not measurable, mainly due to
faintness in the `U' or `B' bands or due to image defects.  This is a sample of the full table
which is available in the online version of the article.}
{\scriptsize
\begin{tabular}{rrrrrrrrrrrr}
\hline
X~~ & Y~~ &$U\;\;\;$&$B\;\;\;$&$V\;\;\;$&$R\;\;\;$&$I\;\;\;$ &$\sigma_{U}\;\;$& $\sigma_{B}\;\;$&$\sigma_{V}\;\;$&$\sigma_{R}\;\;$&$\sigma_{I}\;\;$\\
\hline
126.1 &17.2 &16.025 &14.581 &12.981 &12.113 &99.999 &0.012 &0.002 &0.003 &0.020 &99.999 \\
 67.8 &18.5 &18.052 &17.490 &16.470 &15.902 &15.130 &0.016 &0.006 &0.004 &0.003  &0.003 \\
 91.4 &33.5 &16.783 &16.419 &15.383 &14.752 &13.990 &0.012 &0.001 &0.002 &0.003  &0.001 \\
\hline
\end{tabular}  
}
\end{minipage}
\end{table*}

\begin{table*}
\begin{minipage}{185mm}
\caption{The same as Table 1 but for Be~80.  This is a sample of the full table
which is available in the online version of the article.}
{\scriptsize
\begin{tabular}{rrrrrrrrrrrr}
\hline
X~~ & Y~~ &$U\;\;\;$&$B\;\;\;$&$V\;\;\;$&$R\;\;\;$&$I\;\;\;$&$\sigma_{U}\;\;$& $\sigma_{B}\;\;$&$\sigma_{V}\;\;$&$\sigma_{R}\;\;$&$\sigma_{I}\;\;$\\
\hline
560.3 &19.5 &17.364 &16.915 &15.490 &14.534 &13.453  &0.044  &0.001 &0.002 &0.002 &0.003 \\
806.4 &28.6 &19.120 &18.482 &17.101 &16.204 &15.169  &0.292  &0.022 &0.004 &0.007 &0.003 \\
172.6 &37.1 &99.999 &99.999 &18.994 &17.176 &15.362 &99.999 &99.999 &0.032 &0.012 &0.007 \\
\hline
\end{tabular}  
}
\end{minipage}
\end{table*}

\begin{table*}
\begin{minipage}{185mm}
\caption{The same as Table 1 but for NGC~2192.  This is a sample of the full table
which is available in the online version of the article.}
{\scriptsize
\begin{tabular}{rrrrrrrrrrrr}
\hline
X~~ & Y~~ &$U\;\;\;$&$B\;\;\;$&$V\;\;\;$&$R\;\;\;$&$I\;\;\;$&$\sigma_{U}\;\;$& $\sigma_{B}\;\;$&$\sigma_{V}\;\;$&$\sigma_{R}\;\;$&$\sigma_{I}\;\;$\\
\hline
637.9 &13.7 &18.589 &18.475 &17.739 &17.333 &16.859 &0.024 &0.032 &0.011 &0.008 &0.015 \\
935.5 &20.9 &18.546 &18.569 &17.892 &17.485 &17.018 &0.036 &0.015 &0.023 &0.010 &0.009 \\
640.6 &25.0 &17.145 &17.112 &16.491 &16.176 &15.790 &0.012 &0.006 &0.002 &0.006 &0.005 \\
\hline
\end{tabular}  
}
\end{minipage}
\end{table*}

\section[]{Berkeley 15}

Be~15 is located in the constellation Auriga, in the Galactic anti-centre
direction ($\ell,b = 162.27\degr,+01.61\degr$), has an apparent diameter of about
5.0 arcmin (Lyng{\aa} 1987), and was classified as II-2-p by Trumpler (1930) and
reclassified as I-2-m by Lyng{\aa} (1987), indicating a medium to poor cluster ($\la$
50--100 stars) with a moderate central condensation, and a medium luminosity contrast
with respect to the surrounding fields.  Lyng{\aa} (1987) gives $\approx 15.0$ mag as the
visual magnitude of its brightest star, and this cluster has also been designated as
OCl~414 and C~0458+443 in Simbad.

\begin{figure}
 \epsfig{file=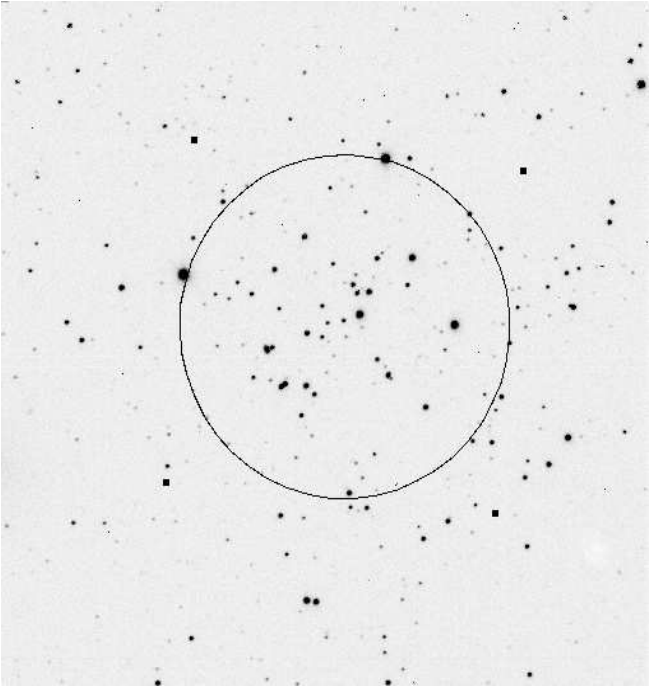, width=8.6cm, height=9.1cm}
\caption{Visual image of the open cluster Be~15 as seen on the DPSS.  The nearly circular
region encloses those stars studied in this publication; details are given in the text.
North is up, and east to the left in this image}
\end{figure}

\begin{figure}
 \epsfig{file=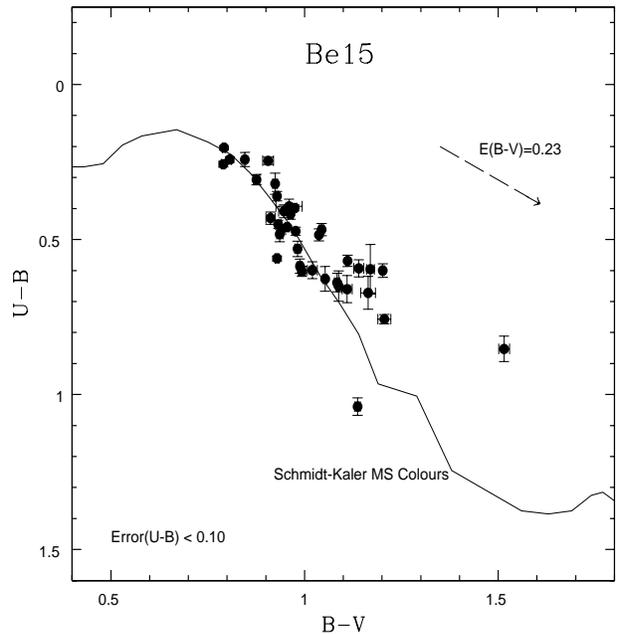, width=8.6cm, height=9.1cm}
 \caption{The CC, $(U$--$B)$ vs $(B$--$V)$, figure for the open cluster Be~15.
This cluster has been fit to the main-sequence colours of Schmidt-Kaler (1982), the thick
solid line which crosses the diagram from upper left to lower right.  Only stars with
photometric errors less than $0.10$ mag for the colour $(U$--$B)$ have been plotted.
Photometric error bars, and the reddening vector corresponding to $E(B$--$V) = +0.23$ are
also shown}
\end{figure}

\begin{figure}
 \epsfig{file=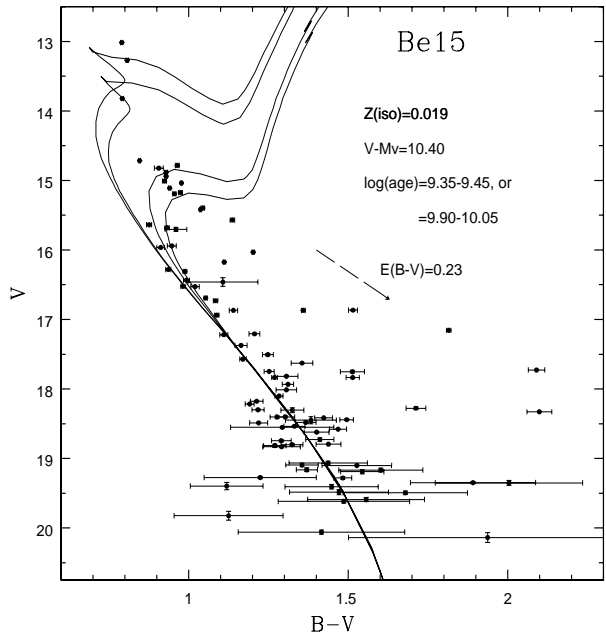, width=8.6cm, height=9.1cm}
 \caption{The CM figure, ($V, B$--$V$), for the open cluster Be~15.  This
cluster has been fit to the isochrones of Girardi et al.~(2000) according to the interstellar
reddening determined above in the CC diagram of Fig.~2, $E(B$--$V)= +0.23$; $[Fe/H] = 0.00$
dex has been assumed.  The distance modulus of Be~15 has been determined by fitting
vertically the cluster to the isochrones at the intermediate magnitudes of the main sequence,
$16.5 \la V \la 18.5$ mag, and the age by fitting the turn-off of the cluster at the brighter
magnitudes, $13.0 \la V \la 16.0$; two possible ages solutions are suggested depending on the
membership of three brighter stars.  Photometric error bars, and the reddening vector
corresponding to $E(B$--$V) = +0.23$ mag are also shown}
\end{figure}

\begin{figure}
 \epsfig{file=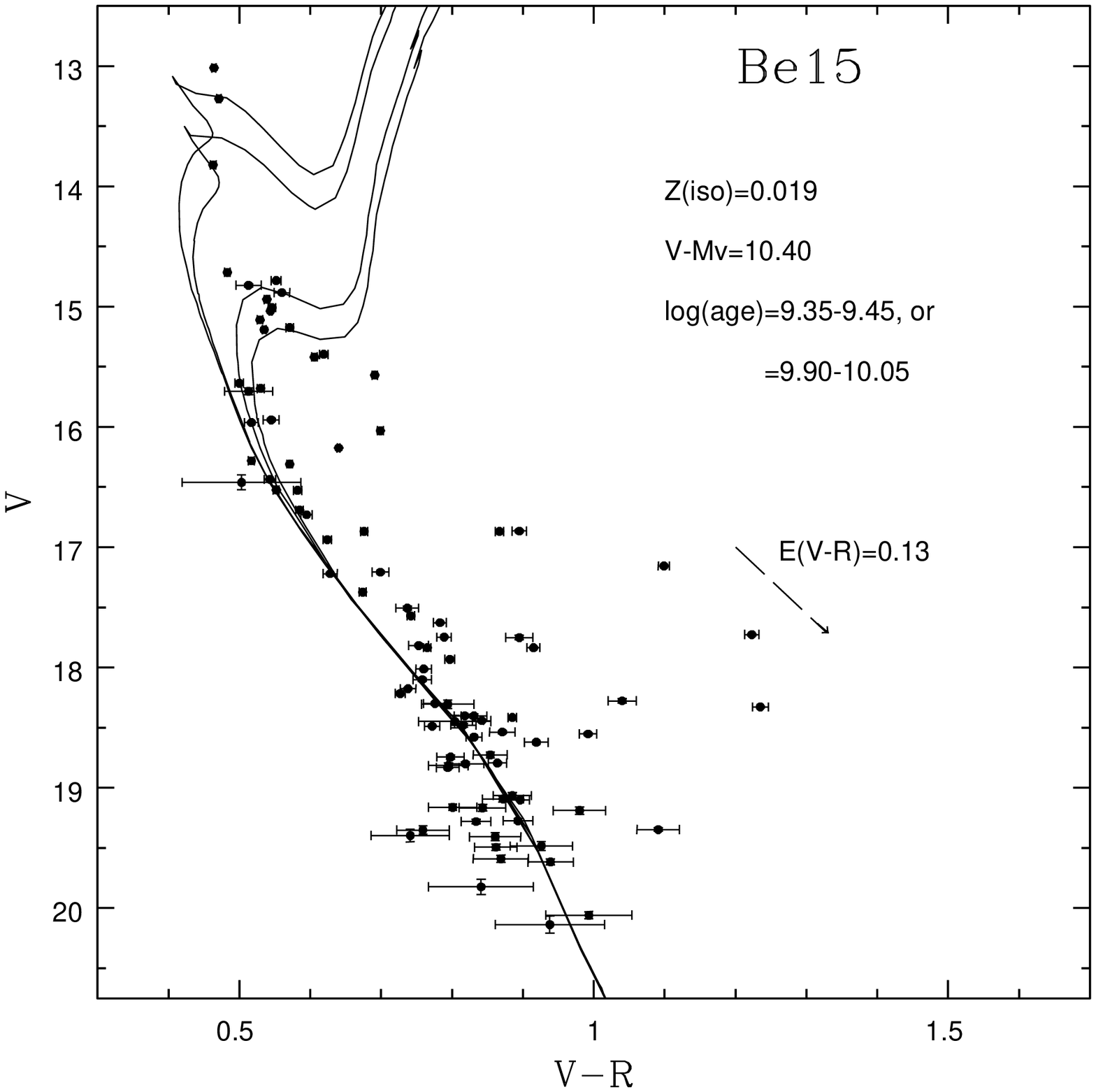, width=8.6cm, height=9.1cm}
 \caption{The CM figure, ($V, V$--$R$), for the open cluster Be~15.  This
cluster has been fit to the isochrones of Girardi et al.~(2000) according to the interstellar
reddening determined above in the CC diagram of Fig.~2, $E(B$--$V)= +0.23$ mag;
$[Fe/H] = 0.00$ dex has been assumed.  Other details as in Fig.~3}
\end{figure}

For the cluster Be~15, the macro `elipse' has been used to retain a nearly circular
region for the photometric analyses (see Fig.~1):  the centre at (487, 482) pixels;
a north-south axis of 256 pixels (1.68 arcmin); an east-west axis of 259 pixels
(1.70 arcmin); and a position angle of $+85\degr$.  The total field of view of
the SITe1 CCD at the 0.84-m telescope is $6.72 \times 6.72$ square arcmin.

\subsection{Colour--colour diagram, $(U$--$B)$ versus $(B$--$V)$}

The $(U$--$B)$ versus $(B$--$V)$ diagram of this open cluster is presented in Fig.~2
together with the Schmidt-Kaler (1982) main-sequence colours, which have been fit assuming
G-type stars, providing an interstellar reddening of $E(B$--$V)= 0.23\pm0.03$ mag for
Be~15.  Only stars with photometric errors less than $0.10$ mag for the $(U$--$B)$ colour
have been plotted to compensate for the low sensitivity of this CCD in the ultraviolet.
The observed slope of these stars in Fig.~2 fits much better the Schmidt-Kaler
G-type main-sequence colours than they do the B-type main-sequence colours, which are
more vertical.  Another study, mentioned below, has fit the observed stars to these
B-type colours.  Photometric error bars, and the reddening vector corresponding to
$E(B$--$V) = +0.23$ are also shown in Fig.~2.  At the redder colours the points begin to
scatter more due to increasing photometric errors in $(U$--$B)$ and $(B$--$V)$,
possible binary stars, and also due to the probable inclusion of background stars with
greater interstellar reddening.  Since this is an older cluster lacking F-type stars, no
means for estimating an ultraviolet excess, $ \delta(U$--$B)$ or $\delta_{0.6}$, is
available, and so a solar metallicity, $Z = 0.019$, has been assumed for Be~15.

Lata et al.~(2004) also studied Be~15 in this same $(U$--$B)$ versus $(B$--$V)$ diagram
from CCD $U\!BV\!RI$ photometry, but they fitted to the B-type main-sequence stars of
Schmidt-Kaler (1982) obtaining a reddening of $E(B$--$V) = +0.88$, much larger than our
result.  In support of our interstellar reddening estimate, the interstellar extinction maps
of Neckel \& Klare (1980) indicate $A_{V} \approx 0.80$ mag, according to their Figs~5b and
6c, for the Galactic longitude and latitude of Be~15, and for the distance derived in our
next Section~4.2 ($d \approx 1.2$ kpc).  This visual absorption corresponds to
$E(B$--$V) \approx +0.28$ mag, much closer to our result than to that of Lata et al.~(2004).
In addition, the reddening value of Be~15 has been derived from the dust maps of Schlegel,
Finkbeiner \& Davis (1998; hereafter, SFD), which are based on the $COBE/DIRBE$ and
$IRAS/ISSA$ maps and take into account the dust absorption all the way to infinity.
A E(B--V)($\ell, b$)$_\infty$ value of $1.22$ for Be~15 has been taken from SFD maps
using the web pages of NED\footnote[2] {http://nedwww.ipac.caltech.edu/forms/calculator.html}.
This value to infinity has been reduced slightly according to the corrections discussed in
Arce \& Goodman (1999), Bonifacio, Monai \& Beers (2000), and Schuster et al.~(2004).
Then the final reddening for a given star is reduced again by a factor
$\lbrace1-\exp[-d \sin |b|/H]\rbrace$, where $b$, $d$, and $H$ are the Galactic latitude,
the distance from the observer to the object, and the scale height of the dust layer in
the Galaxy, respectively; $H = 125$ pc has been assumed here (Bonifacio, Monai \& Beers 2000).
For Be~15 this estimated interstellar reddening works out as $E(B$--$V) = +0.24$, again much
closer to our value than to that of Lata et al.~(2004).  However, the 3~kpc distance of Lata
et al.~(2004) is also much larger than ours leading to larger $A_{V}$ and $E(B$--$V)$ values
from Neckel \& Klare and from SFD, $\approx 2.5$ mag, i.e. $E(B$--$V) \approx +0.81$ mag
(good agreement with Lata et al.'s result), and $0.49$ mag (not so good agreement),
respectively.  So, we prefer our small reddening solution due to the better fit to the
slope of the Schmidt-Kaler colours in Fig.~2, and also due to more consistent fits to the
results from {\bf both} Neckel \& Klare and SFD.

Sujatha et al.~(2004) have also studied Be~15 using $U\!BV\!RI$ CCD photometry, and have
employed the $(B$--$I)$ vs $(B$--$V)$ diagram plus the technique given by Natali et
al.~(1994) obtaining an interstellar reddening of $E(B$--$V)= 0.462$ mag, larger
than our value.  But, their distance modulus and distance are in good agreement with
our estimates:  $10.5$ mag and $1259 \pm 135$ pc (Sujatha et al.~2004) compared to $10.4$
mag and $1202 \pm 50$ pc (our values, next section), providing confidence in our
smaller reddening estimate from Fig.~2, and also in the smaller estimates from Neckel \&
Klare and SFD, which depend on this distance.

\subsection{Colour--magnitude diagrams}

In Figs~3 and 4 are shown the CM diagrams, ($V, B$--$V$) and ($V, V$--$R$),
respectively, employed for Be~15 to determine its distance and age.  In these CM
diagrams the observations have been fit to the isochrones of Girardi et al.~(2000)
using the interstellar reddening determined above, $E(B$--$V)= 0.23$ mag, which
converts to $A_v = 0.71$ mag, assuming $R_{V} = 3.1$, and $E(V$--$R)= 0.13$ mag,
according to the reddening ratios found in Strai\c{z}ys (1995),
$E(V$--$R)= 0.56E(B$--$V)$.  Solar metallicity has been assumed.

For both these CM diagrams the distance moduli have been determined by shifting the
reddening-corrected isochrones vertically to fit the observed $V$ magnitudes at
intermediate values, $16.5 \la V \la 185$ mag.  At these intermediate magnitudes the
cluster main sequence is more inclined (i.e.~less vertical) than at the brighter
magnitudes, but still with fairly small photometric errores, allowing an accurate
fit.  Both diagrams give a distance modulus of $(V$--$M_{\rm V})_{o} = 10.4$ mag
with an estimated uncertainty of about $\pm0.1$ mag.  Likewise, both diagrams have been
used to estimate the cluster ages by fitting the cluster turn-offs to the isochrones
at the brighter magnitudes, $13.0 \la V \la 16.0$ mag.  Both CM diagrams indicate a
rather large uncertainty in the cluster age, depending on whether the three brightest
stars, $V \approx 13.5$ mag, are members or not.  For this reason two possible age
solutions are indicated in each CM diagram, both with a pair of isochrones.  In both
cases the younger age solution corresponds to $\log({\rm age}) \cong 9.35$--$9.45$ by
fitting to these three brightest stars, and an older solution to
$\log({\rm age}) \cong 9.90$--$10.05$ by fitting to a clump of apparent turn-off stars
about $1.5$ mag fainter.  The double isochrones plotted help to appreciate the
uncertainties in these (logarithmic) age estimates, $\approx \pm0.05$ dex.  In Figs~3
and 4, photometric error bars, and the reddening vector corresponding to
$E(B$--$V) = +0.23$ mag are also shown.  The results from these two CM diagrams concerning
the distance modulus and age are very consistent.  A definitive age for Be~15 from our
CCD photometry must await radial velocity and/or proper motion studies to determine
the membership, or not, of the brightest stars in the field of Be~15.

As mentioned above, the results of Sujatha et al.~(2004) agree very well with ours.  They
obtain a distance modulus of $10.5$ mag, and $\log({\rm age}) \approx 9.7$.  On the other hand,
Lata et al.~(2004) obtain a large-distance/small-age solution for Be~15:
$(V$--$M_{\rm V})_{o} = 12.4$ mag and $\log(\rm {age}) = 8.5$, since they fit Be~15 to the
B-type main-sequence stars of Schmidt-Kaler (1982) in the CC diagram.  To discern between our
solution for Be~15 and that of Lata et al.~(2004) will require spectral types for the
brighter stars, to determine whether they are G- or B-types.  The methodology of
Sujatha et al.~(2004) employs the more independent $(B$--$I)$ vs $(B$--$V)$ technique given
by Natali et al.~(1994) for determing the interstellar reddening, and so provides confidence
in our results for a small-reddening/small-distance/large-age solution for Be~15.

On the other hand, most recently Maciejewski \& Niedzielski (2007) have also studied this
cluster, but using only two-filter ($BV$), wide-field CCD photometry with only a single
CM diagram used to derive all the cluster parameters (distance modulus, reddening, and age)
by means of $\chi^2$ fitting to the solar-metallicity isochrones of Bertelli et al.~(1994).
It is obvious in the Be~15 panel of their Fig.~2 that their decontamination of the field of
Be~15 has been only partially successful, due to the remaining presence of field red-giant
stars and also anomalous stars which extend beyond the main sequence of this cluster.
Maciejewski \& Niedzielski obtain a large-reddening/large-distance/small-age solution for
Be~15, similar to that of Lata et al.~(2004):  $E(B$--$V)= 1.01\pm0.15$ mag,
$(V$--$M_{\rm V})_{o} = 12.15\pm0.4$ mag, and $\log({\rm age}) = 8.7\pm0.1$.

\section[]{Berkeley 80}

\begin{figure}
 \epsfig{file=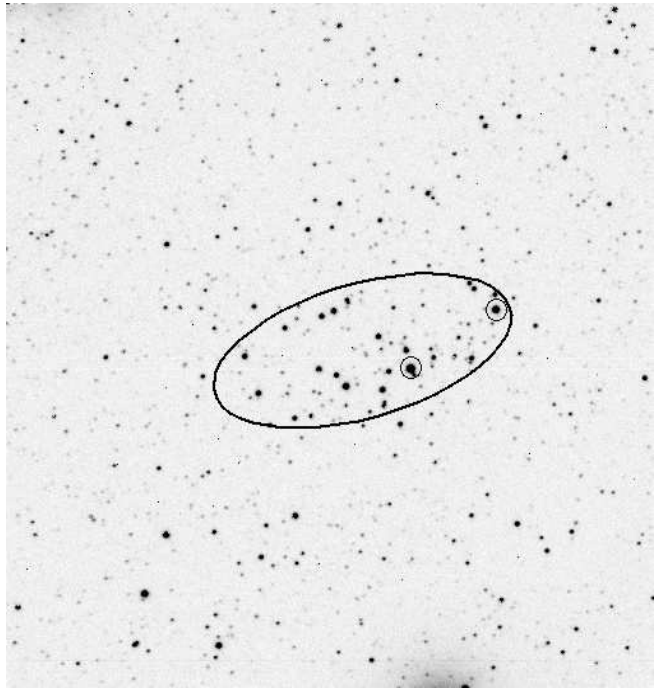, width=8.6cm, height=9.1cm}
\caption{Visual image of the open cluster Be~80 from the DPSS.  The ellipse encloses the
region studied in this publication, and the two small circles mark the brightest two
stars in this elliptical region.  The second brightest is most surely a foreground star
as seen in Figs~6, 8, and 9, while the bightest is that unusual star far to the right in
Fig.~6.  North is up, and east to the left in this presentation}
\end{figure}

This open cluster is located in the constellation Aquila, in the Galactic central
region ($\ell,b = 32.18\degr,-01.21\degr$), has an apparent diameter of 3--4 arcmin
(Dias et al.~2002; Lyng{\aa} 1987), was classified as II-2-p by Trumpler (1930)
and reclassified as II-1-p by Lyng{\aa} (1987), indicating a poor cluster ($\la 50$
stars) with a little central condensation, and a narrow to medium luminosity contrast
with respect to the surrounding fields.  Lyng{\aa} (1987) gives $\approx 15.0$ mag as the
visual magnitude of the brightest star.  This cluster has also been designated as
C~1851-013 in Simbad.  Be~80 is about $2.5\degr$ from the Serpens Cauda dark band of
the Milky Way, and contains the infrared source $IRAS$18518-0117 at the very southern
edge of our elliptical study area (cf.~Fig.~5); this IR source is the brightest object of our I
images of Be~80, fourth brightest in the $R$ images of the study area, quite faint in the
visual ($V \approx 17.45$ mag), and essentially undetected in $B$ and $U$.  To improve the
contrast and visibility of this cluster in the various CC and CM diagrams an elongated
ellipse containing the central region of Be~80 has been defined with the macro `elipse'
(see Fig.~5); this cluster appears very prolonged in all of its images:  centre at
(520,460) pixels, north-south axis 110 pixels (0.72 arcmin), east-west axis 240
pixels (1.58 arcmin), and a rotation of $105\degr$, from north toward east.

\begin{figure}
 \epsfig{file=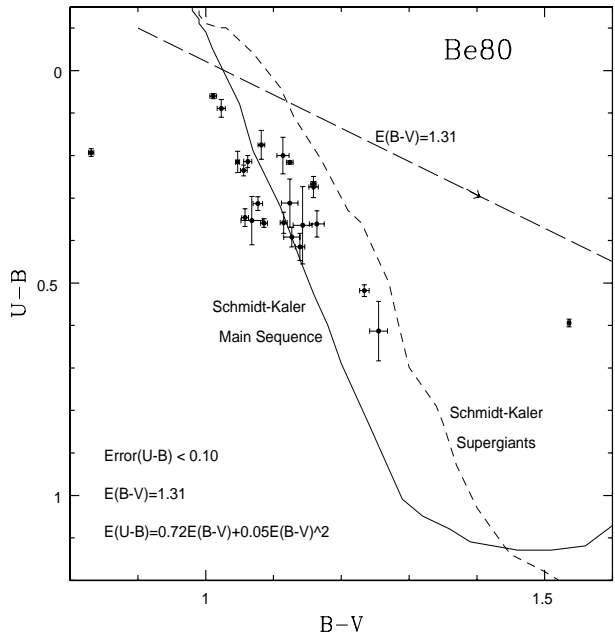, width=8.6cm, height=9.1cm}
 \caption{The CC, $(U$--$B)$ vs $(B$--$V)$, figure for the open cluster Be~80.
This cluster has been fit to the main-sequence colours of Schmidt-Kaler (1982), the solid
line from upper left to lower right; the mostly parallel short-dash line represents the
supergiant colours of Schmidt-Kaler.  Only stars with photometric errors less than
$0.10$ mag for the colour $(U-B)$ and within the ellipse defined in the text (Fig.~5) have
been plotted.  Photometric error bars, and the reddening vector corresponding to
$E(B$--$V) = +1.31$ mag are also plotted (long-dashes).  Due to the large interstellar
reddening of Be~80, the second order term of $0.05E(B$--$V)^2$ has been included here
in the reddening equation}
\end{figure}

\begin{figure}
 \epsfig{file=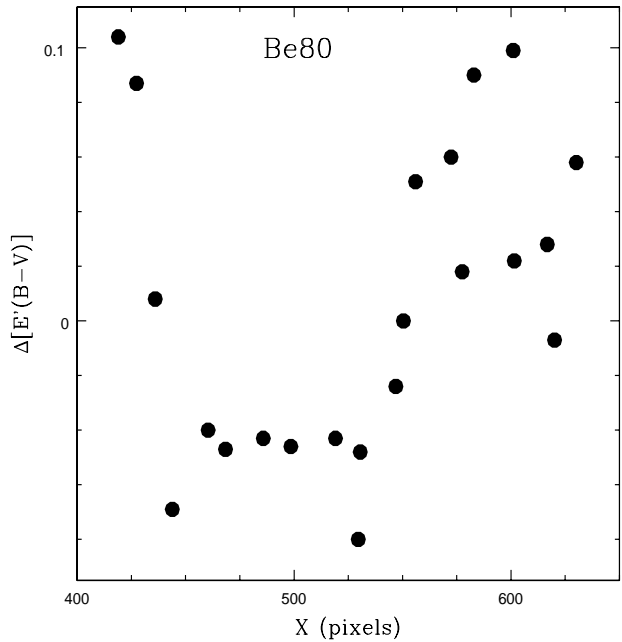, width=8.6cm, height=9.1cm}
 \caption{A measure of the scatter in Fig.~6 is plotted as a function of the star's
north-south position in pixels.  The ordinate of this figure, $\Delta[E'(B$--$V)]$,
is measured along a reddening vector from the Schmidt-Kaler main-sequence relation
for B-type stars to each star's position in the $(U$--$B)$ vs $(B$--$V)$ diagram
of Fig.~6.  Structure indicating variable interstellar extinction for Be~80 is
clearly seen with an obvious minimum for the extinction over pixels 440-550}
\end{figure}

\subsection{Colour--colour diagram, $(U$--$B)$ versus $(B$--$V)$}

The $(U$--$B)$ versus $(B$--$V)$ diagram of this open cluster is presented in Fig.~6
together with the Schmidt-Kaler (1982) main-sequence colours, which has been fit to the
B-type stars providing an interstellar reddening of $E(B$--$V)= 1.31\pm0.05$ mag for
Be~80.  Only stars with photometric errors less than $0.10$ mag for the colour $(U$--$B)$
have been plotted to compensate for the low sensitivity of this CCD in the ultraviolet,
and only stars within the ellipse of Fig.~5 have been included.  Photometric error bars,
and the reddening vector corresponding to $E(B$--$V) = +1.31$ mag are also shown in Fig.~6.
Due to the large interstellar reddening of this cluster, the reddening fit of Fig.~6
has been done with the equation $E(U$--$B) = 0.72E(B$--$V) + 0.05E(B$--$V)^2$, including
the second-order term, which increases the final reddening value by about $0.03$ mag.  The
points of Fig.~6 show considerable scatter about the Schmidt-Kaler main-sequence line
suggesting variable interstellar extinction at this region of the sky as discussed in
the next section.

Carraro et al.~(2005) have also studied Be~80 using CCD $BVI$ photometry and have estimated
a reddening of $E(B$--$V)= 1.10\pm0.05$ mag by matching several CM diagrams with the same
isochrones of Girardi et al.~(2000).  They apparently adjusted simultaneously for the
reddening, distance modulus, and age in these CM diagrams, assuming also a solar metallicity.
Since they did not observe Be~80 in the ultraviolet, they could not make use of the
$(U$--$B, B$--$V)$ diagram as we have done here.  Using Figs 5a and $6\ell$ from Neckel
\& Klare (1980) and a distance of 1.3--1.4 kpc (from Section~5.3 below,
$V$--$M_{\rm V} \cong 10.75$ mag), $A_{\rm V}(r) \cong 4.0$ mag is estimated from their reddening
maps, which agrees satisfactorily with our value of $A_{\rm V} = 3.1E(B$--$V) = 4.06$ mag.

In Fig.~5 the two brightest stars of the ellipse are indicated with small circles.
Star No.~1 is the brightest, falls at the lower centre of the ellipse, and far to the
right in Fig.~6.  Star No.~2 is the second brightest, is located at the right edge of
the ellipse, and far to the left in Fig.~6.  Star 2 is a foreground star with a smaller
distance and less interstellar reddening than the cluster.  The brightest, Star 1, is
more difficult to explain; its position to the right in this CC diagram would suggest
more interstellar reddening, but its brightness would suggest a foreground star.

\subsection{Probable variable interstellar extinction }

The reddening solution for Be~80 has been obtained by fitting the Schmidt-Kaler (1982)
main-sequence colours to those 22 stars falling within the ellipse of this cluster's
approximate shape (Fig.~5) and having the smaller photometric observing errors in the
colour $(U$--$B)$ ($\sigma_{(U-B)} < 0.10$ mag), by assuming that these stars are B-type
stars, and by excluding the two extreme stars mentioned above.  This fit has given a
large value for the interstellar reddening ($E(B$--$V) = 1.31$ mag), but the stars show a
rather large scatter, $\approx 0.20$ mag, about the Schmidt-Kaler line in this CC diagram
(Fig.~6), suggesting variable interstellar, or intracluster, extinction across this
object.

\begin{figure}
 \epsfig{file=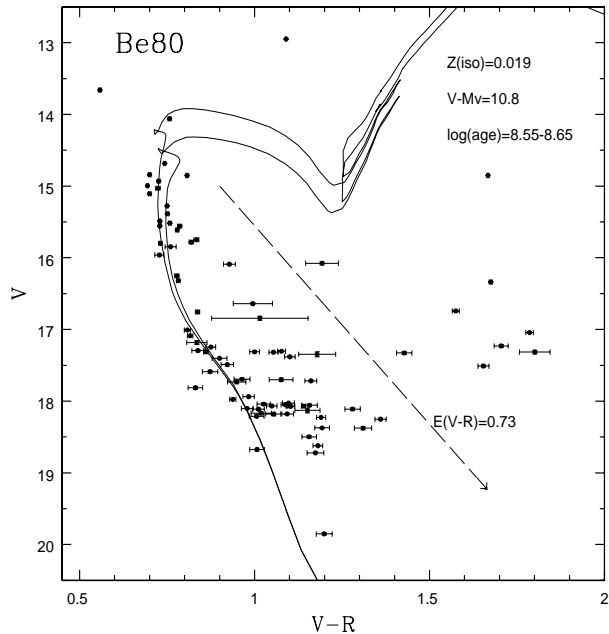, width=8.6cm, height=9.1cm}
 \caption{The CM figure, ($V, V$--$R$), for the open cluster Be~80, which
has been fit to the isochrones of Girardi et al.~(2000) according to the interstellar
reddening $E(B$--$V)= +1.31$ mag and a solar metallicity.  The distance modulus has been
determined by fitting vertically to the isochrones at the intermediate magnitudes of
the main sequence, $16.5 \la V \la 18.0$ mag, and the age by fitting the turn-off at the
brighter magnitudes, $14.0 \la V \la 16.0$.  Photometric error bars, and the reddening
vector corresponding to $E(B$--$V) = +1.31$ mag are also shown}
\end{figure}

\begin{figure}
 \epsfig{file=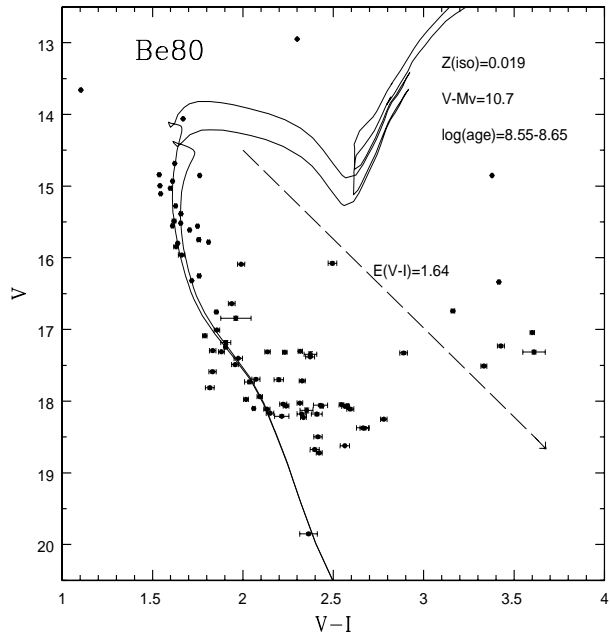, width=8.6cm, height=9.1cm}
 \caption{The CM figure, ($V, V$--$I$), for the open cluster Be~80.  This
cluster has been fit to the isochrones of Girardi et al.~(2000) using $E(B$--$V)= 1.31$ mag and
a metallicity of $[Fe/H] = 0.00$ dex.  Other details as in Fig.~8
}
\end{figure}

In Fig.~7 the scatter of this CC diagram, measured along a reddening vector, is
plotted as a function of the north-south (or X-axis) postion of a star in pixels.
Structure is clearly seen with a minimum in the reddening for a band of pixels
from about 440 to 550; this 0.72 arcmin band runs approximately east-west.
To the north-west three stars have larger reddenings, and to the south-east 10
stars show increasing reddening with increasing pixel value. A plot of this
scatter against Y, the east-west position in pixels, shows no clear structure.
A colour plot against (X, Y) shows that this minimum in reddening runs approximately
east-north-east to west-south-west, slightly skewed, $\approx 30\degr$, with
respect to  to the major axis of the elliptical study area, but many more stars
with precise values of $E(B$--$V)$ are needed for any firm conclusions.  The
amplitude of this scatter in Fig.~7 is $0.184$ mag, corresponding to
$\Delta E(B$--$V) \cong 0.15$ mag, and $\Delta A_v \cong 0.46$ mag.  This structure
suggests the possibility that this cluster is a phantom cluster caused by an
extinction minimum in the interstellar medium allowing us to see more background
stars.  However, the appearance of our CM diagrams for this region (in the next
section) does indicate that there is an open cluster here, and CM diagrams from
2MASS (Skrutskie et al.~2006), such as ($K, K$--$J$), also suggest the presence
of a cluster.  Only much deeper $U\!BV$ CCD data for this region, providing many
more stars with good errors in $(U$--$B)$ for mapping the interstellar extinction
with better extent and spatial resolution, could help us to understand the
variable interstellar absorption of this interesting region.

\subsection{Colour--magnitude diagrams}

In Figs~8 and 9 are shown two of the CM diagrams examined for Be~80 to determine its
distance and age, the ($V, V$--$R)$ and the ($V, V$--$I)$, respectively.  These
redder CM diagrams have been preferred in order to mitigate the effects of the large
and variable interstellar reddening of this cluster.  In these diagrams the CCD
observations of this cluster have been fit to the isochrones of Girardi et al.~(2000)
according to the interstellar reddening determined above $E(B$--$V)= 1.31$ mag, which
converts to $A_v = 4.06$, $E(V$--$R)= 0.73$, and $E(V$--$I)= 1.64$ mag, according
to the reddening ratios of Strai\c{z}ys (1995), $E(V$--$R)= 0.56E(B$--$V)$ and
$E(V$--$I)= 1.25E(B$--$V)$.  Solar metallicity has been assumed, since F-type stars
are not available for measuring an ultraviolet excess.

The distance moduli have been determined by shifting the reddening-corrected isochrones
vertically to fit the observed $V$ magnitudes at intermediate values,
$16.5 \la V \la 18.0$ mag.  The ($V, V$--$R$) diagram provides a distance modulus of
$(V$--$M_{\rm V})_{o} = 10.8$ mag, while the ($V, V$--$I$) gives
$(V$--$M_{\rm V})_{o} = 10.7$, both with uncertainties of $\pm0.1$ mag.  Likewise, both
diagrams have been used to estimate the cluster age by fitting the cluster turn-offs to
the isochrones at the brighter magnitudes, $14.0 \la V \la 16.0$.  Both the
($V, V$--$R$) and ($V, V$--$I$) diagrams give age estimates of
$8.55 \la \log({\rm age}) \la 8.65$.  Again two isochrones are plotted in each of the
diagrams to give a feeling for the uncertainties involved here, and again the results
from these two CM diagrams concerning the distance modulus and age are very consistent.

When comparing our results to those of Carraro et al.~(2005), if we compare our {\bf apparent}
distance modulus of $V$--$M_{\rm V} = 10.75 + A_{\rm V} = 14.81 \pm 0.1$ mag to their
apparent value of $14.8 \pm 0.2$, an almost perfect agreement is found.  So, the only
difference is that between the two values for the interstellar reddening,
$E(B$--$V) = +1.31$ mag (ours) and $+1.10$ (theirs).


\begin{figure}
 \epsfig{file=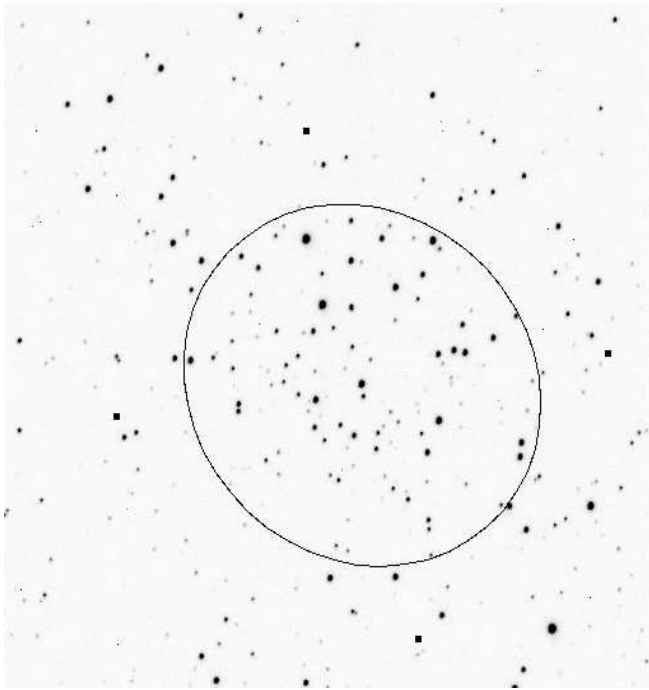, width=8.6cm, height=9.1cm}
\caption{Visual image of the open cluster NGC~2192, as seen on the DPSS.  The ellipse
encloses the region studied in this publication.  North is up and east to the left}
\end{figure}

\begin{figure}
 \epsfig{file=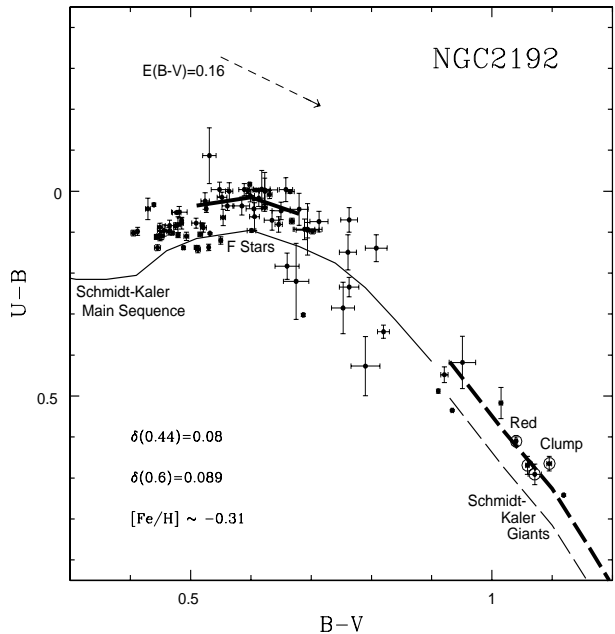, width=8.6cm, height=9.1cm}
 \caption{The CC, $(U$--$B)$ vs $(B$--$V)$, figure for the open cluster NGC~2192.  This
cluster has been fit to the main-sequence colours of Schmidt-Kaler (1982), the solid line
from $(B$--$V) = +0.30$ to $+0.90$ mag, assuming an ultraviolet excess, $\delta(0.44) = 0.08$ mag,
for the F-type cluster stars, represented by the thick solid line over $0.51\le(B$--$V)\le0.68$ mag;
this fit gives $E(B$--$V) = +0.16\pm0.03$ for the interstellar reddening.  At this same reddening
the red-clump stars fit very well the red-giant (III) colours of Schmidt-Kaler, the dashed
line over $(B$--$V)\ge0.93$ mag, with a corresponding ultraviolet excess given by the thick dashed
line.  The reddening vector corresponding to $E(B$--$V) = +0.16$ mag is also shown}
\end{figure}

\section[]{NGC~2192}

This open cluster is located in the constellation Auriga, in the Galactic anti-centre
direction ($\ell,b = 173.42\degr,+10.65\degr$), has an apparent diameter of about
5.0 arcmin (Dias et al.~2002) and was classified as III-1-p by Trumpler (1930) and
reclassified as II-2-m by Lyng{\aa} (1987), indicating a medium to poor cluster ($\la$
50--100 stars) with a little or no central condensation, and a narrow to medium luminosity
contrast with respect to the surrounding fields.  Lyng{\aa} (1987) gives 45 as the number
of member stars, $\approx 14.0$ mag as the visual magnitude of the brightest star, and $10.7$
mag as the total visual magnitude of this cluster.  This cluster has also been designated
C~0611+398 and Cl~Melotte~42 in Simbad.  To concentrate on the central region in order
to increase contrast with respect to the surrounding field, the awk macro `elipse'
has been applied to extract those stars centred at (570,460) pixels with an elliptical
form of 260 pixels north-south (1.71 arcmin), 290 pixels east-west (1.90 arcmin),
and a rotation of $+55\degr$, from north toward east (see Fig.~10).

\begin{figure}
 \epsfig{file=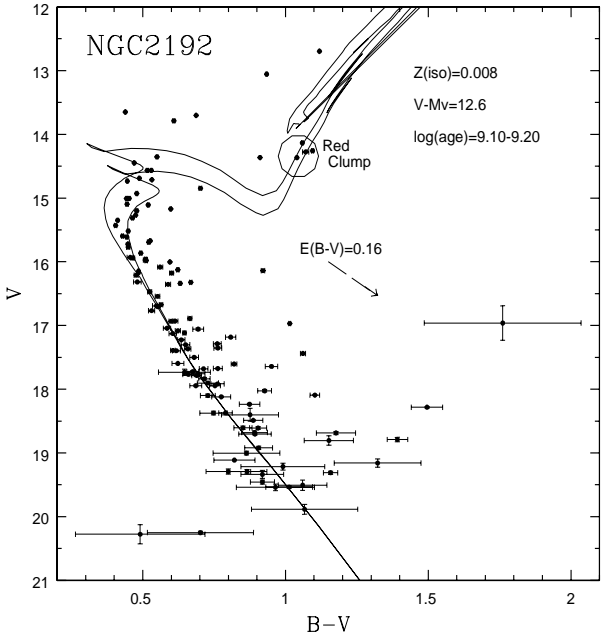, width=8.6cm, height=9.1cm}
 \caption{The CM figure, ($V, B$--$V$), for the open cluster NGC~2192.  This
cluster has been fit to the isochrones of Girardi et al.~(2000) according to $E(B$--$V)= +0.16$
mag and $[Fe/H] = -0.31$ dex.  The distance modulus has been determined by shifting vertically the
cluster to the isochrones at the intermediate magnitudes of the main sequence,
$17.0 \la V \la 18.5$ mag, and the age by fitting the turn-off of the cluster at the brighter
magnitudes, $14.5 \la V \la 16.5$.  Photometric error bars, the position of the red-clump
stars, and the reddening vector corresponding to $E(B$--$V) = 0.16$ mag are also shown}
\end{figure}

\begin{figure}
 \epsfig{file=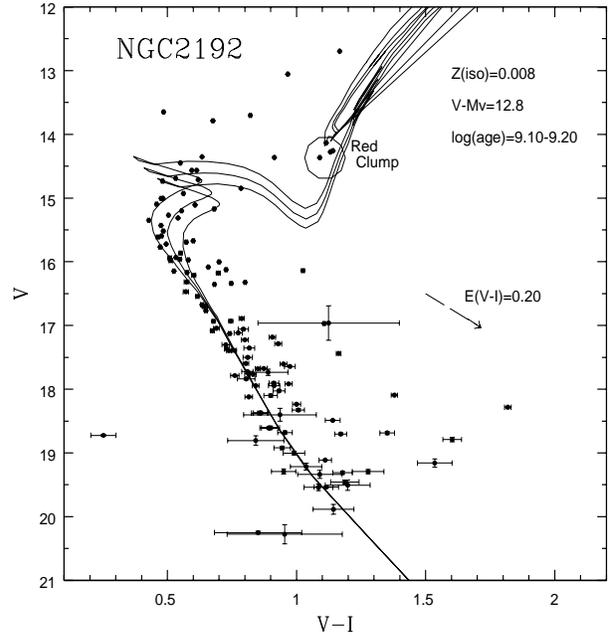, width=8.6cm, height=9.1cm}
 \caption{The CM figure, ($V, V$--$I$), for the open cluster NGC~2192.  This
cluster has been fit to the isochrones of Girardi et al.~(2000) according to $E(B$--$V)= +0.16$
mag and $[Fe/H] = -0.31$ dex, derived above.  Other details as in Fig.~12}
\end{figure}

NGC~2192 is one of 72 open clusters having the most accurately determined parameters, selected
by Paunzen \& Netopil (2006) from a statistical analisis of 395 open clusters found in the
literature with a total of 6437 estimates for their parameters.  These 72 clusters serve as a
standard list for future comparisons and tests between different isochrones and stellar models.
The interstellar reddening, distance, and age derived for NGC~2192 by Paunzen \& Netopil agree
excellently with those of Park \& Lee (1999), well within observational errors.

\subsection{Colour--colour diagram, $(U$--$B)$ versus $(B$--$V)$}

The $(U$--$B)$ versus $(B$--$V)$ diagram of this open cluster allows a nicely consistent fit
for both the interstellar reddening and the metallicity using simultaneously the F-type
stars and the red-clump stars (cf.~Fig.~11).  The Schmidt-Kaler (1982) main-sequence
colours are fit to the F-type stars with the same interstellar reddening,
$E(B$--$V)= 0.16\pm0.03$ mag, as used to fit the red-clump stars to the Schmidt-Kaler giant
colours.  However, there is one complication in that the best fit between the F-type stars
and the Schmidt-Kaler main sequence occurs for a slight ultraviolet excess of
$\delta(0.44) = +0.08$ mag, but this value agrees very well with the resulting ultraviolet
excess at the red-clump stars when compared to the Schmidt-Kaler giant colours for this
same interstellar reddening.  The resulting red-clump ultraviolet excess of
$\delta(B$--$V) \approx +0.09$ mag agrees very well with that seen at the F-type stars
according to the normalization of these excesses as a function of $B$--$V$,
given in Table~1A of Sandage (1969).  The top of the F-star hump for the Schmidt-Kaler
$(U$--$B)$ vs $(B$--$V)$ relation corresponds to $(B$--$V) = +0.44$ mag (unreddened), while the
red-clump stars to $(B$--$V) \approx +0.90$ mag (unreddened).  According to Table~1A of Sandage
both excesses correct to $\delta(0.6) \approx +0.089$ mag at $(B$--$V) = +0.60$ mag, which is that
$(B$--$V)$ value at which most $[Fe/H]$ calibrations of the $U\!BV\!(RI)_C$ photometric system
have been made.  From equation (8) of Karata\c{s} \& Schuster (2006), $\delta(0.6) = +0.089$ mag
gives $[Fe/H] = -0.31\pm0.05$ dex for NGC~2192, which agrees excellently with the $[Fe/H]$
derived from the $\delta(0.6),[Fe/H]$ calibration of Sandage \& Fouts (1987), also
$[Fe/H] = -0.31$.  Park \& Lee (1999) in their CCD $U\!BV\!I$ study of NGC~2192 have also obtained
$[Fe/H] = -0.31\pm0.15$ dex, making use of this same $(U$--$B)$ vs $(B$--$V)$ diagram.

\subsection{Colour--magnitude diagrams}

Figs 12 and 13 show two of the CM diagrams examined to determine the distance and age
for NGC~2192, the ($V, B$--$V$) and the ($V, V$--$I$), respectively.  In these
diagrams the observations have been fit to the isochrones of Girardi et al.~(2000)
according to the $E(B$--$V)= +0.16$ mag and $[Fe/H] = -0.31$ dex ($Z = 0.008$)
obtained in Section~6.1.  $E(B$--$V)= +0.16$ mag converts to $A_v = 0.496$ mag and
$E(V$--$I)= +0.20$ mag, according to the reddening ratios found in Strai\c{z}ys (1995).

For both CM diagrams of Figs.~12 and 13, the distance moduli have been determined by
shifting the reddening-corrected isochrones vertically until they fit the observed $V$
magnitudes at intermediate values, $17.0 \la V \la 18.5$ mag.  The ($V, B$--$V$) and ($V,
V$--$I$) diagrams give distance moduli of $(V$--$M_{\rm V})_{o} = 12.6$ mag and $12.8$,
respectively.  Likewise, both CM diagrams have been used to estimate the cluster ages by
fitting the cluster turn-offs to the isochrones at the brighter magnitudes,
$14.5 \la V \la 16.5$.  Both diagrams indicate a cluster age in the range,
$9.10 \la log({\rm age}) \la 9.20$ as shown by the multiple isochrones plotted:
$\log({\rm age}) = 9.10, 9.15$, and $9.20$; these multiple isochrones also help to estimate the
uncertainties in these (logarithmic) age estimates, $\approx \pm0.05$ dex.  These
ages of $\log({\rm age}) \approx 9.10$--9.20 also provide a fairly good fit to the red-clump
stars, as seen in both figures.  Also, our values for the distance modulus and age of NGC~2192
agree very well, within the estimated observational errors, with those derived by Park
\& Lee (1999):  $12.7 \pm0.2$ mag and $9.04\pm0.06$, respectively.

Maciejewski \& Niedzielski (2007) have also studied NGC~2192 using wide-field $BV$ CCD photometry
and only a single CM diagram ($V, B$--$V$) to obtain the interstellar reddening, distance
modulus, and age using a $\chi^2$ fitting to the solar metallicity isochrones of Bertelli et
al.~(1994). In the NGC~2192 panel of their Fig.~2, it is clear that their decontamination
procedure has worked much better for this cluster.  Due to the differences in metaliicity and
procedure, they have obtained cluster parameters different from those of this paper and of Park
\& Lee (1999):  $E(B$--$V)= 0.04\pm0.13$ mag, $(V$--$M_{\rm V})_{o} = 12.0\pm0.5$ mag, and
$\log({\rm age}) = 9.3\pm0.1$.

\section[]{Discussion and conclusions}

In Table~4 are summarized the results from this study.  When the San Pedro M\'artir open cluster
survey began, only the cluster NGC~2192 had previous studies (Park \& Lee 1999), but in the
meantime three studies for Be~15 (Sujatha et al.~2004; Lata et al.~2004; Maciejewski \& Niedzielski
2007) have appeared as well as one for Be~80 (Carraro et al.~2005), and another for NGC~2192
(Maciejewski \& Niedzielski 2007).  Our results for NGC~2192 agree well with those from Park
\& Lee for all parameters:  reddening, metallicity, distance, and age, as can be seen in Table~4;
but we disagree somewhat with Maciejewski \& Niedzielski (2007) due to a difference in the assumed
metallicity (they assume $Z = 0.019$) and due to their use of only a single CM diagram to derive all
parameters, i.e.~no CC diagram to estimate the interstellar reddening.  For Be~15 our results
agree fairly well with those of Sujatha et al.~(2004) despite a difference in our derived reddening.
Our results do not agree with those from Lata et al.~(2004) and Maciejewski \& Niedzielski (2007)
due mainly to the fact that they fit the main sequence to early-type, main-sequence colours, while
we have fit to the G-type, main-sequence colours of Schmidt-Kaler (1982).  For Be~80 our
{\bf apparent} distance modulus and age agree very well with those from Carraro et al.~(2005), but
our distances do not agree, mainly due to differing values for the interstellar reddening.  They
have fit isochrones of Girardi et al.~(2000) to several CM diagrams from $BVI$ photometry to
estimate simultaneously the distance modulus, reddening, and age, while we have used the
$(U$--$B, B$--$V)$ diagram to separate the reddening solution from that for the distance modulus
and age, providing a more independent and reliable result.

\begin{table*}
\tiny
\begin{minipage}{175mm}
\caption{The inferred fundamental parameters of the three open clusters.
The cluster name and Galactic coordinates are presented in Columns~1 and 2,
respectively.  The derived reddening $E(B$--$V)$ together with its uncertainty is
given in Column~3, and the metallicity and heavy-element abundances, $[Fe/H]$ and
($Z$), in Columns~4 and 5, respectively.  The average values of the
distance modulus, ($V$--$M_{\rm V})_{o}$, heliocentric distance (pc), and
$\log({\rm age})$ (age in years), together with their uncertainties, are listed
in Columns~6, 7, and 8, respectively.  The corresponding references from the
literature are shown in Column~9.}
{\scriptsize

\begin{tabular}{lclcclcll}
\hline
Cluster & $(\ell^{\circ},~b^{\circ})$ & $E(B$--$V)$ (mag)& $[Fe/H]$ (dex) & $Z$ & (V-M$_{\rm V})_{o}$ (mag) & $d$(pc) & $\log({\rm age})$(age in yrs) & Reference \\
\hline
\multicolumn{ 1}{l}{{Be~15}}     &$162.33,~+1.61$&$0.23\pm0.03$&$ 0.00$&+0.019&$10.4 \pm0.1$&1202&$9.35$, or 9.95$\pm0.05$&This paper\\
\multicolumn{ 1}{l}{{}}          &               &$0.462$        &   $-$ &solar &$10.5         $&1259&$9.7            $&Sujatha et al.~(2004)\\
\multicolumn{ 1}{l}{{}}          &               &$0.88\pm0.05$&   $-$ &solar &$12.4 \pm0.2$&3051&$8.5\pm0.1      $&Lata et al.~(2004)\\
\multicolumn{ 1}{l}{{}}          &               &$1.01\pm0.15$&   $-$ &solar &$12.15 \pm0.4$&2690&$8.7\pm0.1      $&Maciejewski \&...(2007)\\
\hline
\multicolumn{ 1}{l}{{Be~80}}     &$ 32.17,~+1.25$&$1.31\pm0.05$&$ 0.00$&+0.019&$10.75\pm0.1$&1413&$8.6\pm0.05     $&This paper\\
\multicolumn{ 1}{l}{{}}          &               &$1.10\pm0.05$&   $-$ &solar &$11.39\pm0.2$&1897&$8.5\pm0.15     $&Carraro et al.~(2005)\\
\hline
\multicolumn{ 1}{l}{{NGC~2192}}  &$173.41,+10.64$&$0.16\pm0.03$&$-0.31$&+0.008&$12.7 \pm0.1$&3467&$9.15\pm0.05    $&This paper\\
\multicolumn{ 1}{l}{{}}          &               &$0.20\pm0.03$&$-0.31$&+0.009&$12.7 \pm0.2$&3467&$9.04\pm0.06    $&Park \& Lee (1999)\\
\multicolumn{ 1}{l}{{}}          &               &$0.04\pm0.13$&    $-$ &solar &$12.0 \pm0.5$&2500&$9.3\pm0.1      $&Maciejewski \&...(2007)\\
\hline
\end{tabular}
}
\end{minipage}
\end{table*}

\subsection{Conclusions}

\begin{enumerate}

\item{Our best values for the interstellar reddening, distance modulus, and $\log({\rm age})$
for Be~15 are $E(B$--$V)= 0.23\pm0.03$ mag, $(V$--$M_{\rm V})_{o} = 10.4\pm0.1$ mag, and 
$\log({\rm age}) \cong 9.35\pm0.05$ mag or $9.95\pm0.05$, respectively.  The two possible
solutions for the cluster age of Be~15 depend upon the membership, or no, of the three
brightest stars, as seen in Figs~3 and 4.}

\item{Our results for Be~15 agree well with those of Sujatha et al.~(2004), who employ the
more independent $(B$--$I)$ vs $(B$--$V)$ technique given by Natali et al.~(1994) for
determing the interstellar reddening, but do not agree well with the results of Lata et
al.~(2004) nor with those of Maciejewski \& Niedzielski (2007); the former have fit the
main-sequence stars of Be~15 to the B-type, main-sequence colours of Schmidt-Kaler (1982),
while the latter to the early-type isochrones of Bertelli et al.~(1994); we have fit Be~15
to the G-type, main-sequence colours of Schmidt-Kaler.  This suggests the need for
spectroscopy for the brighter stars of Be~15 to determine the membership of the
brightest stars using radial velocity measurements, and spectral-type classifications
to know whether the brighter members are in fact G-types or B-types.}

\item{Our best values for the interstellar reddening, distance modulus, and $\log({\rm age})$
for Be~80 are $E(B$--$V)= 1.31\pm0.05$ mag, $(V$--$M_{\rm V})_{o} = 10.75\pm0.1$ mag, and 
$\log({\rm age}) \cong 8.60\pm0.05$, respectively.  Our value for the reddening is in
good agreement with that derived from the extinction maps of Neckel \& Klare (1980).  Our
{\bf apparent} distance modulus and age agree very well with the values of Carraro, the
main difference between our solutions is the interstellar reddening; they obtain
$E(B$--$V)= 1.10\pm0.05$ mag.}

\item{Clear evidence for variability of the interstellar extinction across the face of
the open cluster Be~80 has been detected.  The maximum of this change occurs approximately
in the north-south direction, and the reddening has a minimum with a width of about 0.72
arcmin, running approximately east-north-east to west-south-west, slightly north of the
cluster's centre, and slightly skewed, $\approx 30\degr$, with respect to  to the major
axis of our elliptical study area.  Much deeper CCD $U\!BV\!(RI)_C$ photometry is
recommended to map more thoroughly the interstellar extinction in the vicinity of Be~80
and to study the structure of the interstellar, or intracluster, reddening of this
interesting region.}

\item{A very consistent and simultaneous solution for the interstellar reddening and
metallicity of NGC~2192, $E(B$--$V) = +0.16\pm0.03$ mag and $[Fe/H] = -0.31\pm0.05$, is
obtained from the CC $(U$--$B, B$--$V)$ diagram using both the F-type and the red-clump
stars.  Park \& Lee (1999) obtained this identical metallicity by means of a similar study
using CCD $U\!BV\!I$ photometry; they obtained a slightly larger reddening of
$E(B$--$V) = +0.20\pm0.03$ mag.  Our distance modulus for NGC~2192,
$(V$--$M_{\rm V})_{o} = 12.7\pm0.1$ mag, from Figs~12 and 13, agrees very well with that
from Park \& Lee (1999),  $(V$--$M_{\rm V})_{o} = 12.7\pm0.2$ mag.  The ages also agree
within the quoted errors, $\log({\rm age}) = 9.15\pm0.05$ from this paper, and
$\log({\rm age}) = 9.04\pm0.06$ from Park \& Lee.}

\item{Our results for NGC~2192 do not agree as well with those from Maciejewski \& Niedzielski
(2007), but they use only a single CM diagram ($V, B$--$V$) to estimate all cluster
parameters, i.e. no CC diagram for the interstellar reddening, and assume a solar metallicity
for the isochrones.}

\end{enumerate}

\section*{Acknowledgments}  We wish to thank the staff of the San Pedro M\'artir Observatory
for their help with the CCD observations, and also we greatly appreciated the participation of
C.A.~Santos, and J.~McFarland, who assisted with the data reductions and programming.
This research made use of the WEBDA open cluster database of J.-C. Mermilliod.  This work was
supported by the CONACyT projects 33940, 45014, 49434 and PAPIIT-UNAM IN111500 (M\'exico).

\section*{SUPPORTING INFORMATION}

Additional Supporting Information may be found in the online version of this article:\\
Table~1. Standard $U\!BV\!RI$ CCD photometry and observing errors for the open cluster Be~15.\\
Table~2. The same as Table~1 but for Be~80.\\
Table~3. The same as Table~1 but for NGC~2192.\\

\label{lastpage}

\end{document}